\documentclass[pra,twocolumn]{revtex4-1}

\usepackage{amsmath}
\usepackage{amsfonts}
\usepackage{tikz}
\usepackage{graphicx}
\usepackage{subfigure}
\usepackage{bm}
\usepackage{color}

\newcommand{\ket}[1]{\ensuremath {|\: #1 \: \rangle}}
\newcommand{\bra}[1]{\ensuremath{\langle \: #1 \:|}}

\newcommand{\ketbra}[2]{\ensuremath{| \: #1 \:\rangle \langle \: #2 \:  |}}

\newcommand{\eref}[1]{(\ref{#1})}
\newcommand{\fref}[1]{figure \ref{#1}}
\newcommand{\Fref}[1]{Figure \ref{#1}}

\newcommand{\llrr}[1]{\ensuremath{\left( #1\right)}}
\newcommand{\llrrq}[1]{\ensuremath{\left[ #1\right]}}

\begin{document}

\title{Time-dependent density functional theory for open spin chains}


\author{Diego de Falco}
\author{Dario Tamascelli}
%\thanks{}
\affiliation{Dipartimento di Scienze dell'Informazione, Universit\`a degli Studi di Milano\\
Via Comelico, 39/41, 20135 Milano- Italy}
\email{defalco@dsi.unimi.it,tamascelli@dsi.unimi.it}


\date{\today}

\begin{abstract}
The application of methods of time-dependent density functional theory (TDDFT) to systems of qubits provided the interesting possibility of simulating an assigned Hamiltonian evolution by means of an auxiliary Hamiltonian having different two-qubit interactions and hence a possibly simpler wave function evolution. In this note we extend these methods to some instances of Lindblad evolution of a spin chain.
\end{abstract}

%03.67.Lx Quantum computer architectures and implementations.
%03.67.Ac
%\keywords{}

\maketitle

\section{Introduction \label{sec:intro}}
\begin{figure}[!h]
\centering
\includegraphics[width=1.\columnwidth]{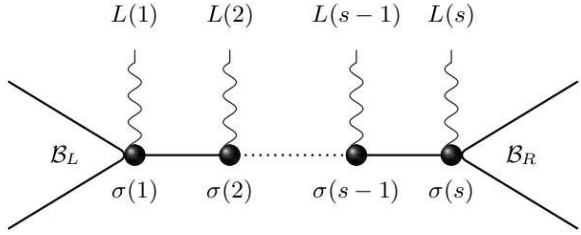}
\caption{The end points of a spin chain of length $s$ interact with two reservoirs $\mathcal{B}_L$, $\mathcal{B}_R$. Dephasing operators act independently on each site.\label{fig:unit}}
\end{figure}

The application of methods of time-dependent density functional theory to systems of qubits has been proposed by Aspuru-Guzik and Tempel in reference \cite{aspuru11}. In particular, these authors have given, for a family of Hamiltonian evolutions of qubit systems, a constructive proof of the van Leeuwen theorem (VL) \cite{vanLeeu99}.
\\We review their work on the Hamiltonian evolution of a spin chain in section II: we explore the possibility offered by van Leeuwen's mapping from densities to potentials of simulating, by means of a spin chain having spatially homogeneous hopping parameters, the magnetization of an \emph{engineered} spin chain \cite{christandl04}.
\\Section III explores the possibility of extending these methods to some instances of dissipative evolution of a spin chain according to the Lindblad equation. We focus our attention on a class of models studied in all details by \v{Z}nidari\v{c} \cite{znidaric10bis}\cite{znidaric11} and examplified in \fref{fig:unit}.
\\[5pt]The case in which only dephasing is present is of some interest on its own right: in subsection III.A we numerically explore the possibility (suggested in \cite{tempel10} for electronic systems) of simulating, in this case, the magnetization of an \emph{open} quantum system with that of a driven \emph{closed} system; then we discuss the complementary problem of compensating for the effects of dephasing on a spin chain. Section III.B is devoted to the study of the dissipation current \cite{gebauer04} \cite{tempel11} present when we allow exchange of magnetization with the two reservoirs $\mathcal{B}_L,\mathcal{B}_R$.
\\[5pt]In section IV we discuss the difficulties, already evidenced in \cite{tokatly11}, that appear when the local kinetic energy is not everywhere nonzero, and present suggestions for further work.

%%%%%%%%%%%%%%%%%%%%%%%%%%%%%%%%%%%%%%%%%

%%%%%%%%%%%%%%%%%%%%%%%%%%%%%%%%%%%%%%%%%
\section{The models: Hamiltonian evolution}\label{sec:models}
We consider a chain of $s$ spin 1/2 systems $\bm{\sigma}(1),\bm{\sigma}(2),\ldots,\bm{\sigma}(s)$, with $\bm{\sigma}(x)=\llrr{\sigma_1(x),\sigma_2(x),\sigma_3(x)}$, evolving under the action of a Hamiltonian of the form:
\begin{align} \label{eq:ham}
 & H(t) = \sum_{x=1}^{s-1} J(x) (\sigma_1(x)\sigma_1(x+1) + \sigma_2(x)\sigma_2(x+1))+\nonumber \\
 &+\sum_{x=1}^{s-1} K(x) \sigma_3(x)\sigma_3(x+1)+\sum_{x=1}^s h(x,t) \sigma_3(x)= \nonumber \\
 &=H_0(J,K)+ \sum_{x=1}^s h(x,t) \sigma_3(x).
\end{align}
As our notations indicate, the two-qubit coupling constants $J(x)$ and $K(x)$ are supposed to be time independent, while the field $h(x,t)$ is allowed to depend on time.
\\The current operator field $\mathrm{j}$ is defined by:
\begin{align} \label{eq:currentop}
& \mathrm{j}(0)=\mathrm{j}(s)=0 \nonumber\\
& \mathrm{j}(x) = 2 J(x)(\sigma_1(x)\sigma_1(x+1) - \sigma_2(x)\sigma_2(x+1)).
\end{align}
Its interest comes from the following commutator identity (from which the continuity equation easily follows):
\begin{align} \label{eq:continuity1}
 &-i [\sigma_3(x),H(t)]+\llrr{\mathrm{j}(x)-\mathrm{j}(x-1)}=0.
\end{align}
It is useful, in what follows, to consider also the identity:
\begin{align}
 \mathrm{j}(x) = 4i J(x) (\sigma_+(x)\sigma_-(x+1)-\sigma_-(x)\sigma_+(x+1)),
\end{align}
where $\sigma_\pm(x) = \llrr{\sigma_1(x) \pm i \sigma_2(x)}/2$.
\\Using the above identities it is immediate to prove the following further identity, of obvious relevance to the Heisenberg evolution of the current field:
\begin{align} \label{eq:dj}
 &-i [\mathrm{j}(x),H(t)] = \\
 & 8J(x) \left ( J(x)\llrr{\sigma_3(x)-\sigma_3(x+1)}+ \right.\nonumber \\
 & +J(x-1) \sigma_3(x) \tau(x-1,x+1)+  \nonumber \\
 & - J(x+1) \sigma_3(x+1) \tau(x,x+2)+  \nonumber \\
 & + K(x+1) \sigma_3(x+2) \tau(x,x+1)+  \nonumber \\ 
 & -K(x-1) \sigma_3(x-1) \tau(x,x+1)+ \nonumber \\
 & \left . +(h(x+1,t)-h(x,t)) \tau(x,x+1) \right ). \nonumber
\end{align}
Here and in what follows we set
\[
\tau(x,y)= \sigma_+(x)\sigma_-(y)+\sigma_-(x)\sigma_+(y).
\]
\\Identities \eref{eq:continuity1} and \eref{eq:dj} have the obvious consequence that, if the state \ket{\psi(t)} satisfies
the Schr\"odinger equation
\begin{align}
 i \frac{d}{dt} \ket{\psi(t)} = H(t) \ket{\psi(t)},
\end{align}
then its \emph{mean magnetization} field 
\[
m_3(x,t) = \bra{\psi(t)} \sigma_3(x) \ket{\psi(t)}
\] 
satisfies the continuity equation
\begin{align} \label{eq:continuity2}
 \frac{d m_3(x,t)}{dt}+(j(x,t)-j(x-1,t))=0
\end{align}
and its  \emph{mean current} field 
\[
j(x,t) = \bra{\psi(t)} \mathrm{j}(x) \ket{\psi(t)}
\] 
satisfies the equation of motion
%%RIPRENDI DA QUI
%
\begin{align}
 &\frac{d}{dt} j(x,t) = -i \bra{\psi(t)}[\mathrm{j}(x),H_0(J,K)]\ket{\psi(t)}+  \\
 &+ 8 J(x) (h(x+1,t)-h(x,t)) \bra{\psi(t)} \tau(x,x+1) \ket{\psi(t)} .\nonumber 
\end{align}
\\In reference \cite{aspuru11} Tempel and Aspuru-Guzik take the following point of view: suppose we are interested in attaining, in a neighborhood of $t=0$, a preassigned magnetization field $\overline{m}_3(x,t)$; because of \eref{eq:continuity2} and \eref{eq:currentop} this target will be reached if we are able to attain a current field $\overline{j}(x,t)$ given by
\begin{align}
 \overline{j}(x,t) = -\frac{d}{dt} \sum_{y \leq x}\overline{m}_3(x,t);
\end{align}
finding a Hamiltonian $H_0(J,K)+\sum_{x=1}^s h(x,t) \sigma_3(x)$ (where $J$ and $K$ are supposed to be assigned, and $h$ is supposed to be unknown) that determines such a current field $\overline{j}(x,t)$ is, in turn, equivalent to solving the differential-algebraic problem
\begin{align} \label{eq:diffalg}
\begin{cases}
&\frac{d}{dt} \overline{j}(x,t) = -i \bra{\psi(t)}[\mathrm{j}(x),H_0(J,K)]\ket{\psi(t)}+  \\
&8 J(x) (h(x+1,t)-h(x,t)) \bra{\psi(t)} \tau(x,x+1) \ket{\psi(t)}\\
&\\
& i \frac{d}{dt} \ket{\psi(t)} = \llrr{H_0(J,K)+\sum_{x=1}^s h(x,t) \sigma_3(x)}\ket{\psi(t)}
\end{cases}
\end{align}
in the unknown \emph{force field} $h(x+1,t)-h(x,t)$ and in the unknown \emph{wave function} \ket{\psi(t)}. Naturally, \ket{\psi(t)} is required to have the correct initial magnetization $\overline{m}_3(x,0)$ and the correct initial current $\overline{j}(x,0)$. For this formulation of the van Leeuwen theorem \cite{vanLeeu99} as a non linear Schr\"odinger equation in which the potential is itself a functional of the wave function, we refer the reader to references \cite{tokatly11} and \cite{maitra10}.
\\In the applications, $H_0(J,K)$ typically describes the two-qubit interaction one is physically able to achieve. Suppose however that the task one has to undertake is the transfer of magnetization along the chain \cite{benjamin03, *benjamin04} \cite{vinet11} realizable through a different two-qubit Hamiltonian $H_0(\overline{J},\overline{K})$. The question is: can a \emph{force field} $h(x+1,t)-h(x,t)$ compensate for the ``wrong'' choice of two-body couplings?
\\[5pt]As an example, consider as a \emph{target} system a spin chain evolving under the \emph{engineered XY-Hamiltonian}:
\begin{align} \label{eq:hDatta}
 \overline{H} = -\frac{1}{2} \sum_{x=1}^{s-1} \frac{\pi \sqrt{x(s-x)}}{s} \tau(x,x+1).
\end{align}
The Hamiltonian \eref{eq:hDatta}, first discussed in \cite{peres85} and then rechristened \emph{engineered XY-chain} in \cite{christandl04}, is quite interesting for the quantum computing community: it induces a periodic behavior of the system with period $T=2s$; in particular it realizes the  \emph{perfect transfer} of an excitation (spin up) initially located at the first site $x=1$ to the final site $x=s$ in a time $s$ \cite{vinet11}.
\\\Fref{fig:figHam} summarizes some experience we have gained in the numerical integration of \eref{eq:diffalg} for $s=6$, starting from the initial condition
\begin{align} \label{eq:inCondPeres}
&\ket{\psi_0} = \frac{1}{\sqrt{s}}\sum_{x=1}^s \ket{x},
\end{align}
where we have indicated by 
\[
\ket{x} = \ket{\sigma_3(x)=+1, \sigma_3(y)=-1 \mbox{ for $y \neq x$}}
\]
the simultaneous eigenstate of the operators $\sigma_3$ in which only the spin at position $x$ is ``up ''.
\\As \fref{fig:figHam} shows, without undertaking the difficult task of implementing the engineered couplings, varying as $\sqrt{x\llrr{s-x}}$, it is possible to obtain the same magnetization and current fields using spatially homogeneous two-body couplings and suitably chosen \emph{control fields} $h^c(x,t)$, e.g. by letting, as we did, the initial condition \ket{\psi_0} evolve under
\begin{align} \label{eq:hcontrHam}
 H^c = -\frac{1}{2} \sum_{x=1}^{s-1} \tau(x,x+1) + \sum_{x=1}^s h^c (x,t) \sigma_3(x)  .
\end{align}
%
%%%%%%%%%%%%%%%%%%%%%%%%%%%%%%%%%%%%%%%%%
%OPEN SPIN CHAINS
%%%%%%%%%%%%%%%%%%%%%%%%%%%%%%%%%%%%%%%%%
\section{Open spin chains}
We discuss the possibility of extending the above discussion to the case in which the initial state of the system, described by a density matrix $\rho_0$, evolves according to a Lindblad equation of the form:
\begin{align} \label{eq:Lindblad}
 \frac{d\rho(t)}{dt}=i \left[ \rho(t),H(t)\right]+\mathcal{D}\llrr{\rho(t)},
\end{align}
where $H(t)$ is of the form \eref{eq:ham}.
\\The dissipator $\mathcal{D}$ that we consider takes into account both the dephasing induced by the interaction of each spin with some external degrees of freedom (for example phonons) and the coupling with unequal one-spin baths \cite{znidaric11bis} $\mathcal{B}_L$ and $\mathcal{B}_R$ at both ends of the chain; it has the form: 
\begin{equation} \label{eq:dissForm}
\mathcal{D}(\rho) =  \mathcal{D}^{deph}(\rho)+ \mathcal{D}^{bath}(\rho);
\end{equation}
The dephasing superoperator is given, in Lindblad form \cite{znidaric10bis,znidaric11}, by
\begin{align} \label{eq:dephasing}
 \mathcal{D}^{deph}(\rho) = \sum_{x=1}^s \llrr{\llrrq{L(x)\rho,L(x)^\dagger} + \llrrq{L(x),\rho L(x)^\dagger}},
\end{align}
where the generator associated with each site $x$ is
\[
L(x) = \sqrt{\frac{\eta}{2}} \sigma_3(x).
\]
$\mathcal{D}^{bath}$, instead, involves four generators:
\begin{align}{\label{eq:lindops}}
&L_1 = \sqrt{\epsilon (1-\mu)} \sigma_+(1), &L_2 = \sqrt{\epsilon (1+\mu)} \sigma_-(1),\\
&L_3 = \sqrt{\epsilon (1+\mu)} \sigma_+(s), &L_4 = \sqrt{\epsilon (1-\mu)} \sigma_-(s), \nonumber
\end{align}
where the coupling parameter $\epsilon$ and the asymmetry parameter $\mu$ satisfy the conditions $\epsilon >0$ and $-1 \leq \mu \leq 1$. The asymmetry parameter $\mu$ models a possible difference of the chemical potentials of $\mathcal{B}_L$ and $\mathcal{B}_R$. We refer the reader to the second section of \cite{casati09bis} for a discussion of the range of validity of this approach and of the possible degree of control on the baths and on the bath-system interaction.
\\[5pt]For the sake of clarity, we discuss the effect of the two sources of dissipation separately.
\\[5pt]If there is no interaction with the baths ($\epsilon=0)$, it is, for $1\leq x \leq s$:
\begin{align}
 Tr\llrr{\sigma_3(x) \mathcal{D}(\rho) } = 0.
\end{align}
The continuity equation \eref{eq:continuity1} retains, then, the form
\[
  \frac{d m_3(x,t)}{dt}+(j(x,t)-j(x-1,t))=0
\]
where it is, now, $ m_3(x,t)=Tr\llrr{\sigma_3(x)\rho(t)}$ and \mbox{$j(x,t) = Tr\llrr{\mathrm{j}(x) \rho(t) }$}.
\\Simple algebraic manipulations show that, for $1 \leq x <s$, it is:
\begin{align}
 Tr\llrr{\mathrm{j}(x) \mathcal{D}(\rho)}=-4 \eta Tr\llrr{\mathrm{j}(x) \rho}.
\end{align}
%	\textcolor{red}{(DOMANDA: non e` il caso di citare Kostin qui?)}
The evolution of the current field $j(x,t)$ is modified by the presence of dephasing, namely:
\begin{align} \label{eq:currDeph}
 \frac{d j(x,t)}{dt} &=-i \ Tr \llrr{[\mathrm{j}(x),H(t)]\rho(t) }- 4 \eta j(x,t).
\end{align}
The commutator on the right hand side is given by \eref{eq:dj}.
\\[5pt]The evolution in the presence of the interaction with the baths $\mathcal{B}_L$ and $\mathcal{B}_R$ ($\eta =0,\ \epsilon >0$), on the other side, determines a nontrivial modification of the continuity equation. For $1\leq k \leq 4$, we set
\[
\mathcal{D}L_k\llrr{\rho(t)} = \llrrq{L_k\rho,L_k^\dagger} + \llrrq{L_k,\rho L_k^\dagger}
\]
so that it is 
\begin{align} \label{eq:dissBaths}
\mathcal{D}^{bath}(\rho) = \sum_{k=1}^4 \mathcal{D}L_k(\rho).
\end{align}
A little algebra shows now that the effect of the left bath $\mathcal{B}_L$ on the magnetization is:
\begin{align*}
&Tr\llrr{\sigma_3(x)\llrr{\mathcal{D}L_1(\rho)+ \mathcal{D}L_2(\rho)}}= \\
&= -\delta_{1,x} 4 \epsilon \llrr{\mu+ Tr\llrr{\sigma_3(x)\rho} } \end{align*}
whereas the effect of the right bath $\mathcal{B}_R$ is:
\begin{align*}
&Tr\llrr{\sigma_3(x)\llrr{\mathcal{D}L_3(\rho)+ \mathcal{D}L_4(\rho)}}= \\
&= -\delta_{s,x} 4 \epsilon \llrr{-\mu+ Tr\llrr{\sigma_3(x)\rho} }. \end{align*}
The continuity equation reads now:
\begin{align} \label{eq:continuityBaths}
  &\frac{d m_3(x,t)}{dt}+(j(x,t)-j(x-1,t))= \\
  &=-4 \epsilon \llrr{\delta_{1,x} \llrr{m_3(x,t)+\mu} +\delta_{s,x} \llrr{m_3(x,t)-\mu}}. \nonumber
\end{align}
The rhs of \eref{eq:continuityBaths} plays the role of a \emph{dissipation current} \cite{gebauer04}.\\
As to the time evolution of the current field it is: 
\begin{align}
 \frac{d j(x,t)}{dt} &= -i\ Tr \llrr{[\mathrm{j}(x),H(t)]\rho(t)}+  \nonumber \\ 
 &- 2 \epsilon \llrr{\delta_{1,x}+\delta_{s-1,x}} j(x,t).
\end{align}
\\[5pt]Summarizing, the kinematical and dynamical equations in the presence of both dephasing and coupling with the baths ($\eta >0$ and $\epsilon >0$) are:
\begin{align}
  &\frac{d m_3(x,t)}{dt}+(j(x,t)-j(x-1,t))= \\
  &=-4 \epsilon \llrr{\delta_{1,x} \llrr{m_3(x,t)+\mu} +\delta_{s,x} \llrr{m_3(x,t)-\mu}} \nonumber \\
  &\frac{d j(x,t)}{dt} = -i \ Tr\llrr{[\mathrm{j}(x),H(t)]\rho(t)}+\nonumber \\
  & -\llrr{4 \eta+ \llrr{\delta_{1,x}+\delta_{s-1,x}} 2 \epsilon}  j(x,t). \nonumber
\end{align}
We conclude this section with a few numerical examples.  In all the examples that follow, we take an initial condition of the form $\rho_0=\ketbra{\psi_0}{\psi_0}$, with \ket{\psi_0} given, for different values of $s$, by \eref{eq:inCondPeres}.

\subsection{Pure dephasing}
\begin{figure*}[h]
\centering
$\begin{array}{c}
\subfigure[]{\includegraphics[width=1.\columnwidth]{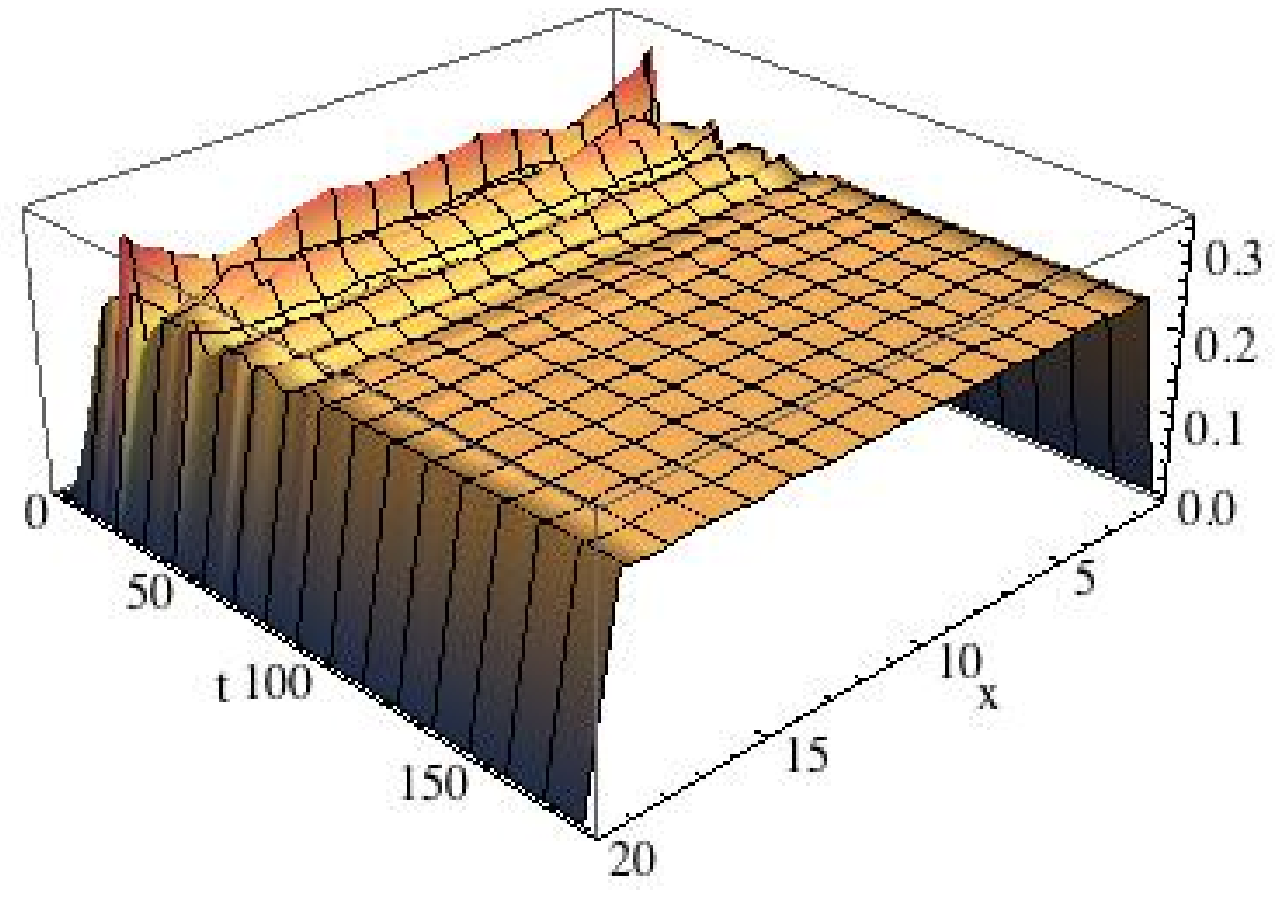}}\hspace{0.1\columnwidth}
\subfigure[]{\includegraphics[width=1.\columnwidth]{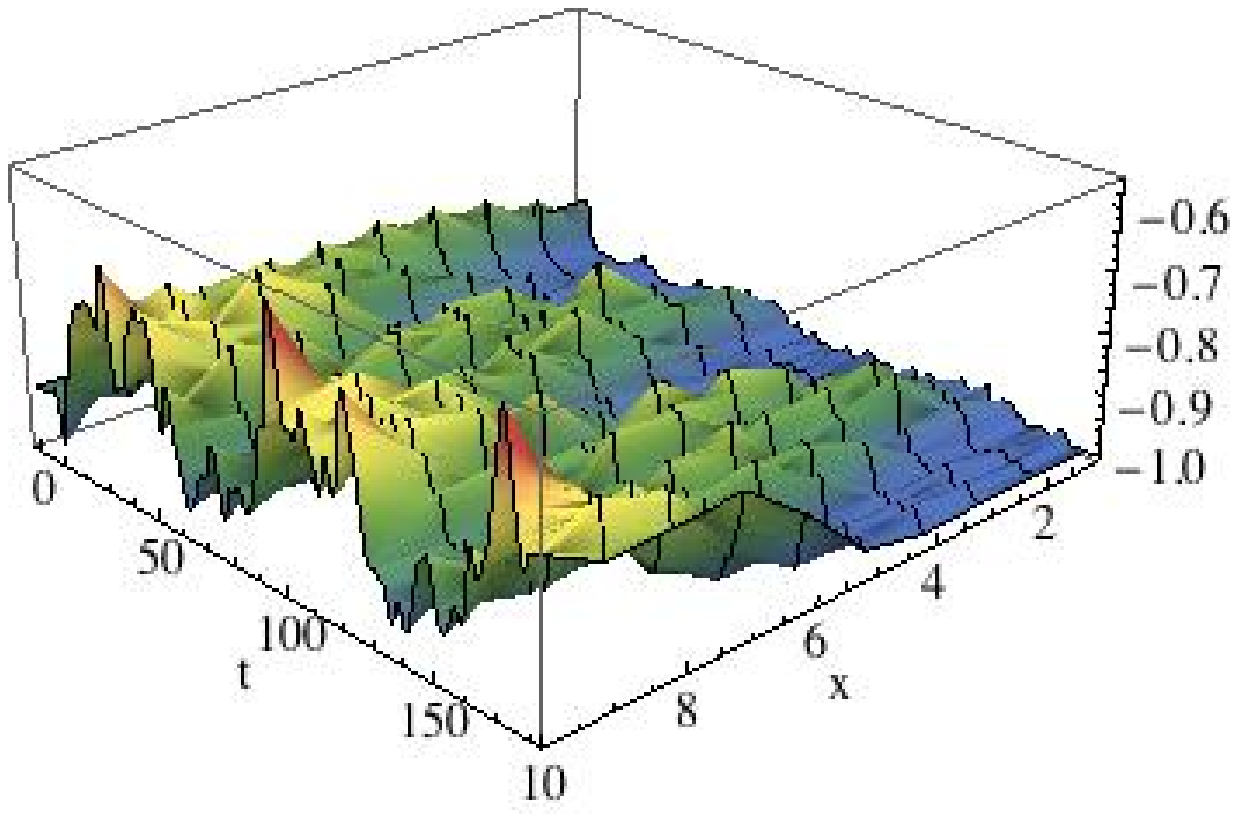}}\\
%\vspace{-0.7cm}\\
\subfigure[]{\includegraphics[width=1.\columnwidth]{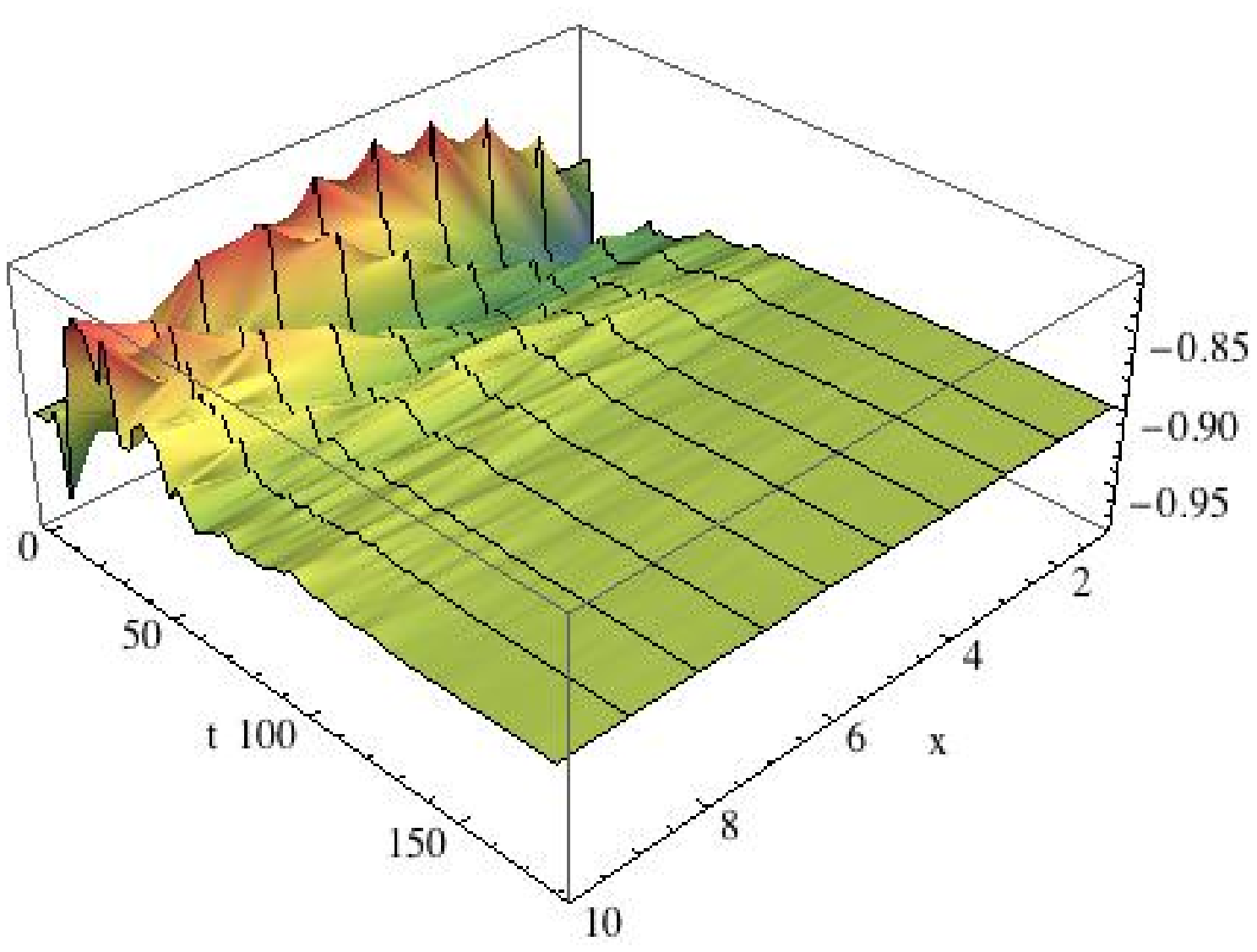}} \hspace{0.1\columnwidth}
\subfigure[]{\includegraphics[width=1.\columnwidth]{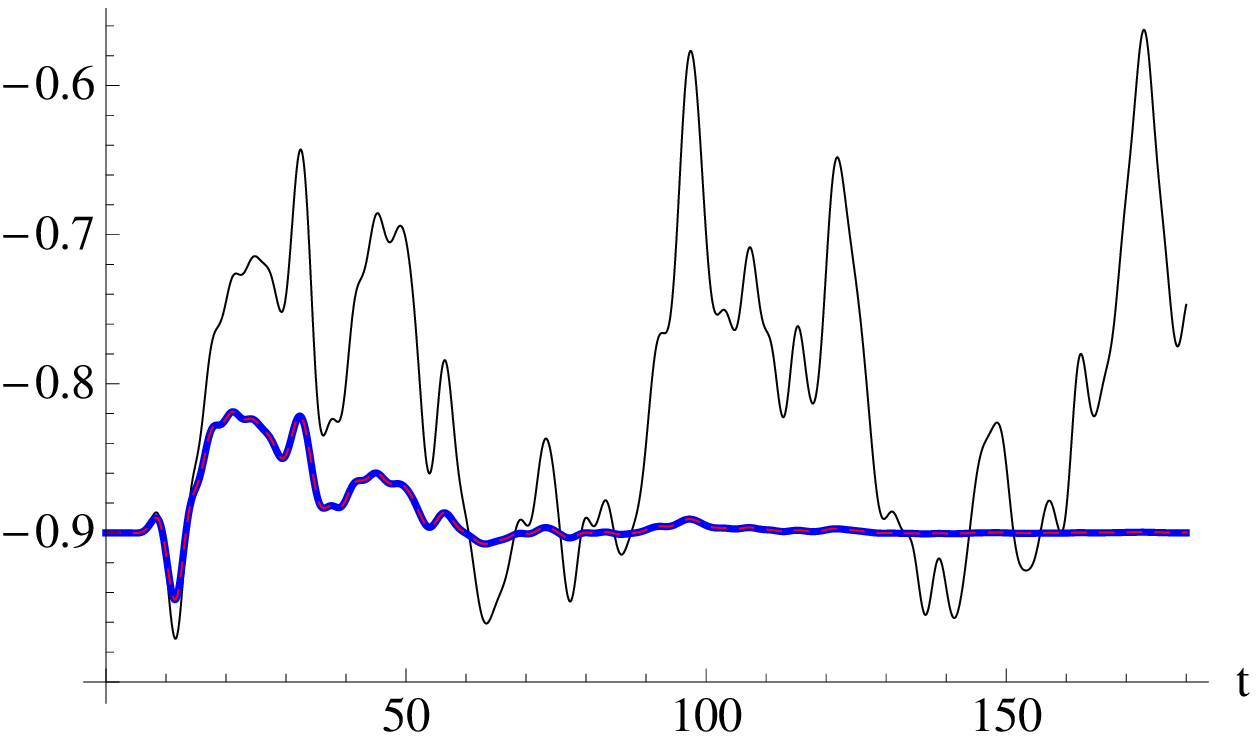}}
\end{array}$
\caption{$s=20,\ \eta=0.01,\ \epsilon=0$. Frame (a): the Kohn-Sham potential $h^c(x,t)$. Frame (b): the field $m_3(x,t)$ corresponding to purely unitary evolution of the initial condition $\rho_0$ under the Hamiltonian \eref{eq:hamZero}. Frame (c): the magnetization field $m_3^c(x,t)$. Frame (d): $m_3(10,t)$ (thin line) and $m_3^c(10,t)$ (thick line). The field $m_3^{deph}(x,t)$ is  numerically indistinguishable from $m_3^c(x,t)$.\label{fig:figRev}}
\end{figure*}

In the case of pure dephasing, the observable
\[
N_3 = \sum_{x=1}^s \frac{1+\sigma_3(x)}{2},
\]
namely the number of \emph{particles} (spins ``up''), is a constant of motion.
\\It has been shown in reference \cite{tempel10} that it is possible to reproduce current and density of an electronic system subject to a number-conserving dissipation by letting the same quantum system evolve unitarily under a time-dependent Hamiltonian with suitably chosen control field $h^c(x,t)$.
\\We wish to show that the same considerations apply to open  spin chains as well.
\\[5pt]Suppose that the interactions in the chain are described by the $XY$-Hamiltonian:
\begin{align} \label{eq:hamZero}
H_0 = -\frac{1}{2}\sum_{x=1}^{s-1}\llrr{ \sigma_+(x) \sigma_-(x+1) +  \sigma_+(x+1)\sigma_-(x)}
\end{align}
that can be obtained from the Hamiltonian family described by \eref{eq:ham} by setting: $J(x)=-1/4,\ K(x)=h(x,t)=0$.\\
The evolution of the chain under the sole effect of dephasing from the initial state $\rho_0$ is determined by the Lindblad equation \eref{eq:Lindblad} with $\mathcal{D} = \mathcal{D}^{deph}$ defined as in \eref{eq:dephasing}. We indicate by $j^{deph}(x,t)$ and $m_3^{deph}(x,t)$ the ensuing current and magnetization fields.\\
We wish to determine an external field $h^c(x,t)$ such that the evolution $\rho^c(t)$ of the initial state $\rho_0$ under the Hamiltonian:
\begin{align}
H^c(t) = H_0+ \sum_{x=1}^s h^c(x,t) \sigma_3(x)
\end{align}
determines the same magnetization and current fields as in the dissipating system, that is:
\begin{align}
 &m_3^c(x,t)  = Tr\llrr{\sigma_3(x) \rho^c(t)} = m_3^{deph}(x,t) \nonumber\\
 &j^c(x,t)= Tr\llrr{\mathrm{j}(x) \rho^c(t)} = j^{deph}(x,t).
\end{align}
It is sufficient to solve, for the state $\rho^c(t)$, the Liouville problem
\begin{align} \label{eq:Liouville}
 &\rho^c(0)=\rho_0 \nonumber\\
 &\frac{d\rho^c(t)}{dt} = i\llrrq{\rho^c(t), H^c(t)}
\end{align}
where the unknown field $h^c(x,t)$ is itself a functional of $\rho^c$ through the algebraic equation:
\begin{align} \label{eq:algdiffRev}
&\frac{d j^{deph}(x,t)}{dt} = -i \ Tr \llrr{[j(x),H_0] \rho^c(t)} + \\
&+ 8 J(x) \llrr{h^c(x+1,t) -h^c(x,t)} Tr\llrr{\tau(x,x+1) \rho^c(t)}. \nonumber
\end{align}
In the numerical example that we discuss below, $d j^{deph}(x,t)/dt$ is \emph{known}, having been determined by numerical integration of the Lindblad equation for pure dephasing under the initial condition $\ketbra{\psi_0}{\psi_0}$ with $s=20$ and $\eta=0.01$.
\\\Fref{fig:figRev} shows that, for this example, the differential-algebraic problem posed by \eref{eq:Liouville} and \eref{eq:algdiffRev} does have a solution that very well attains the current of the dissipating system ($j^c = j^{deph}$) and, because of the continuity equation \eref{eq:continuity1}, also the magnetization ($m_3^c = m_3^{deph}$).
\\The Kohn-Sham potential \cite{tempel10} shown in frame 3.(a) is able to reduce the amplitude of the oscillations visible, for Hamiltonian evolution under $H_0$, in frame 3.(b); this leads to the magnetization profile of frame 3.(c) in which the controlled magnetization is reported: it is numerically indistinguishable from the target magnetization (not reported). Because of the left-right symmetry of our problem we have found it sufficient to let $x$ go only from $1$ to $s/2$.
\\[5pt]As a further example, we pose the following, complementary, problem: we are given an open quantum system affected only by dephasing ($\epsilon=0, \eta>0$) and evolving under a master equation of the form \eref{eq:Lindblad}. Is it possible to determine an external \emph{control} field $h^c$ that compensates for the effects of dissipation, at least as far as the magnetization $m_3$ and the current $j$ fields are concerned? 
\\For the sake of definiteness, we consider a spin chain with $XY$-interaction $H_0$, as in \eref{eq:hamZero}. The system is affected by noise: the dissipator $\mathcal{D}^{deph}$ (see equation \eref{eq:dephasing}) intervenes in the evolution of the system as in \eref{eq:Lindblad}.
\\We are looking for a choice of a control field $h^c$ such that the current and magnetization fields evolve in time as if the dissipator were not acting at all; namely the \emph{target} is:
\begin{align}
 \overline{m}_3(t,x) = Tr\llrr{\sigma_3(x) \overline{\rho}(t)} \nonumber \\
 \overline{j}(x,t) = Tr\llrr{\mathrm{j}(x) \overline{\rho}(t)} \nonumber
\end{align}
where $\overline{\rho}(t)$ is the solution of:
\[
\frac{d\overline{\rho}}{dt} = i \llrrq{H,\overline{\rho}(t)}.
\]
with the initial condition $\overline{\rho}(0)=\rho_0$. 
\\[5pt]\Fref{fig:figDeph} refers to a system of $s=3$ spins. It shows that our control scheme works only \emph{locally}; in fact, as soon as one of the \emph{local kinetic energy} terms $T(x)=2 J(x) \tau(x,x+1)$ hits the value zero, the corresponding control field $h^c$ develops a singularity, as shown in \fref{fig:figDeph}(a). \\It is not surprising that the control problem posed above admits only a solution that is local in time. This aspect of the problem has already been discussed in \cite{tokatly11} and \cite{baer08}.
\subsection{Dissipation current}
In order to discuss the role of equation \eref{eq:continuityBaths}, we analyze here, for the sake of definiteness, the same problem posed at the end of the previous subsection (is there an external field that ``compensates'' for dissipation?) in the presence of interaction with the baths $\mathcal{B}_L,\ \mathcal{B}_R$.
\\The novel feature that appears in the case $\epsilon > 0$ is the fact that the number operator $N_3$ ceases to be a constant of motion: ``charges'' in the form of ``spins up'' can be pumped into or drained from the chain by the baths. Because of this fact, one must carefully distinguish between magnetization and current.
In the discussion that follows, we set
\[
H^c(t) = H_0 +\sum_{x=1}^s h^c(x,t) \sigma_3(x),
\]
with $H_0$ given by equation \eref{eq:hamZero}, and consider the dissipator \eref{eq:dissBaths}
\[
\mathcal{D}^{bath}(\rho) = \sum_{k=1}^4 \mathcal{D}L_k(\rho);
\]
for the definition of the current operator $\mathrm{j}(x)$ we refer to equation \eref{eq:currentop} and, as usual, we take in our numerical example $\rho_0 = \ketbra{\psi_0}{\psi_0}$ with $\ket{\psi_0}$ as in \eref{eq:inCondPeres}.
\\\Fref{fig:figBaths} answers, in the particular case $s=3$, the following question: is there a field $h^c(x,t)$ such that, under the action of a Hamiltonian of the form $H^c(t)$, \underline{and} of the dissipator $\mathcal{D}^{bath}(\rho)$, the initial condition $\rho(0)$ evolves into a state $\rho^c(t)$ for which the expectation value $Tr\llrr{\rho^c(t) \mathrm{j}(x)}$ coincides, for $1 \leq x \leq s-1$ and at least in a neighborhood of $t=0$, with
\[
\overline{j}(x,t) \stackrel{def}{=} \bra{\psi_0} \exp(+itH_0) \mathrm{j}(x) \exp(-itH_0) \ket{\psi_0}?
\]
The behaviour of the control field $h^c(x,t)$ satisfying the condition
\begin{align} \label{eq:conditionj}
Tr\llrr{\rho^c(t) \mathrm{j}(x)} = \overline{j}(x,t)
\end{align}
posed above shows (see \fref{fig:figBaths}(a)) that the existence of the solution is again local, because of the vanishing (\fref{fig:figBaths}(b)) of the expected local kinetic energy $T(1)$ in the controlled evolution.
\\Figures \ref{fig:figBaths}(c) and \ref{fig:figBaths}(d) show that, as long as it is defined, the field $h^c(x,t)$ does enforce the target current: this is hardly surprising since it has been found by imposing precisely \underline{this} condition.
\\In order to compare, in the figures, the controlled evolution with the Lindblad evolution without control, we define $\rho^{actual}(t)$ as the evolution of the initial condition $\rho_0$ determined solely by the Hamiltonian $H_0$ and the dissipator $\mathcal{D}^{bath}$.      
\\[5pt]Setting
\[
\overline{m}_3(x,t)  \stackrel{def}{=} \bra{\psi_0} \exp(+itH_0) \sigma_3(x) \exp(-itH_0) \ket{\psi_0},
\]
\fref{fig:figBathsMag} answers, in the same situation as above, the following different question: for which values of $x$ does the state $\rho^c(t)$ determined by imposing the condition \eref{eq:conditionj} satisfy the additional condition
\begin{align}
 Tr\llrr{\rho^c(t) \sigma_3(x)} = \overline{m}_3(x,t)?
\end{align}
In order to answer this question, we must use equation \eref{eq:continuityBaths}. For $1<x<s$ it is:
\begin{align}
& \frac{d m_3^c(x,t)}{dt} = -\llrr{j^c(x,t) - j^c(x-1,t)} = \nonumber \\
& = -\llrr{\overline{j}(x,t) - \overline{j}(x-1,t)} = \frac{d \overline{m}_3(x,t)}{dt} \nonumber
\end{align}
and
\begin{align}
 m_3^c(x,0) = \overline{m}_3(x,0). \nonumber
\end{align}
Therefore, the controlled magnetization $m_3^c(x,t)$ at sites $2,3,\ldots,s-1$ is, as long as it exists, identical to the target one (\fref{fig:figBathsMag}(a)).
\\As to the sites coupled to the reservoirs, because, again, of equation \eref{eq:continuityBaths}, the following differential equation holds for $m_3^c(1,t)$
\begin{align} 
 & \frac{d m_3^c(1,t)}{dt} = -j^c(1,t) -4 \epsilon \mu - 4 \epsilon m_3^c(1,t) = \nonumber \\
& = -\overline{j}(1,t) -4 \epsilon \mu - 4 \epsilon m_3^c(1,t)  = \nonumber \\
&=\frac{d \overline{m}_3(1,t)}{dt}-4 \epsilon \mu - 4 \epsilon m_3^c(1,t). \nonumber
\end{align}
Under the initial condition $m_3^c(1,0) = \overline{m}_3(1,0)$, this implies that
\begin{align} \label{eq:leftend}
 m_3^c(1,t) &= \overline{m}_3(1,t) - \mu\llrr{1-e^{-4 \epsilon t}} + \nonumber \\
 &- 4 \epsilon e^{-4 \epsilon t} \int_0^t \overline{m}_3(1,\tau) e^{4 \epsilon t} d\tau. 
\end{align}
Similarly, a site $s$, it is
\begin{align}
 m_3^c(s,0) & = \overline{m}_3(s,0) \nonumber \\
 \frac{d m_3^c(s,t)}{dt} &= \frac{d \overline{m}_3(s,t)}{dt}+4 \epsilon \mu - 4 \epsilon m_3^c(s,t) \nonumber
\end{align}
and, therefore
\begin{align} \label{eq:rightend}
 m_3^c(s,t) &= \overline{m}_3(s,t) + \mu\llrr{1-e^{-4 \epsilon t}} + \nonumber \\
 &- 4 \epsilon e^{-4 \epsilon t} \int_0^t \overline{m}_3(s,\tau) e^{4\epsilon t} d\tau. 
\end{align}
Therefore, the controlled magnetizations at sites $1$ and $s$ are, as long as they exist, different from the target ones (see the initial part of \fref{fig:figBathsMag}(b)).
\\[5pt]By explicit computation it is easy to see that, in the simple example $s=3$ at hand, the target magnetizations are given by
\begin{align*} 
\overline{m}_3(2,t) = - \frac{1}{6} \cos\llrr{\sqrt{2} t}
\end{align*} 
and 
\begin{align*} 
\overline{m}_3(1,t) = \overline{m}_3(3,t)= -\frac{1}{2} + \frac{1}{6} \cos\llrr{\sqrt{2} t}.
\end{align*}
With the above explicit expressions of $\overline{m}_3$ and $m_3^c(3,t)$ it is immediate to compute the right hand side of equations \eref{eq:leftend} and \eref{eq:rightend}, namely
\begin{align*}
 a(t)&=\overline{m}_3(1,t) - \mu\llrr{1-e^{-4 \epsilon t}} - 4 \epsilon e^{-4 \epsilon t} \int_0^t \overline{m}_3(1,\tau) e^{4 \epsilon t} d\tau \\
 b(t)&= \overline{m}_3(s,t) + \mu\llrr{1-e^{-4 \epsilon t}} - 4 \epsilon e^{-4 \epsilon t} \int_0^t \overline{m}_3(s,\tau) e^{4\epsilon t} d\tau.
\end{align*}
These are the two quantities represented by the thin dashed and dotted lines in \fref{fig:figBathsMag}(b), extending well beyond the interval \mbox{$0 \leq t \leq 3.55$} of existence of the solution of the VL problem.
\\The fact that there is a value of $t$ beyond which $a(t)$ and $b(t)$ can go out of interval $[-1,1]$ of physically acceptable values of magnetization, convinces us of the fact that the determination of a control field compensating for the dissipation induced by the interaction with the baths $\mathcal{B}_L$ and $\mathcal{B}_R$ admits a solution that is \emph{at most local} in $t$. The local character of the solution is therefore determined neither by the choice of the initial condition, nor by the Hamiltonian and dissipator, nor by the numerical algorithm of solution of the differential algebraic problem.
%
%
%%%%%%%%%%%%%%%%%%%%%%%%%%%%%%%%%%%%%%%%%
\section{Conclusions and outlook}
\begin{figure*}[h]
\centering
\vspace{-1cm}
$\begin{array}{c@{\hspace{1in}}c}
\subfigure[]{\includegraphics[width=1.05\columnwidth]{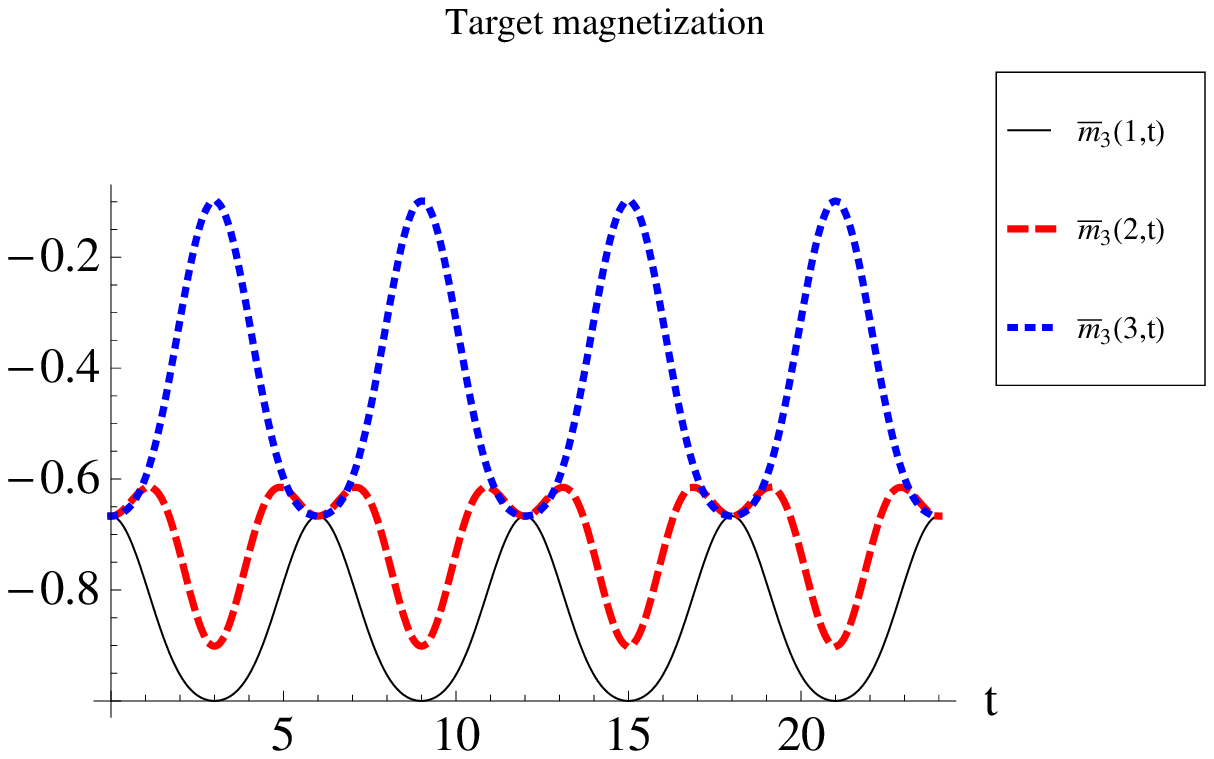}}\hspace{0.\columnwidth}
\subfigure[]{\includegraphics[width=1.05\columnwidth]{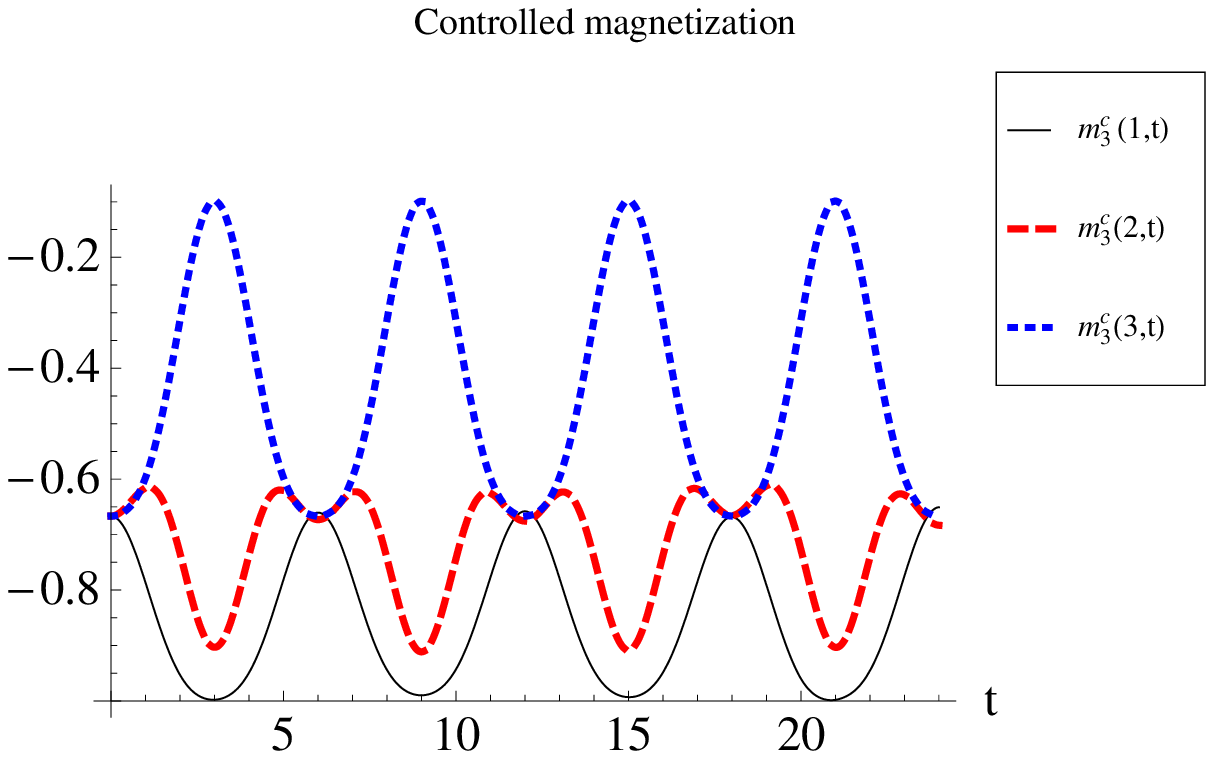}}
\vspace{-0.1cm}\\
\subfigure[]{\includegraphics[width=0.9\columnwidth]{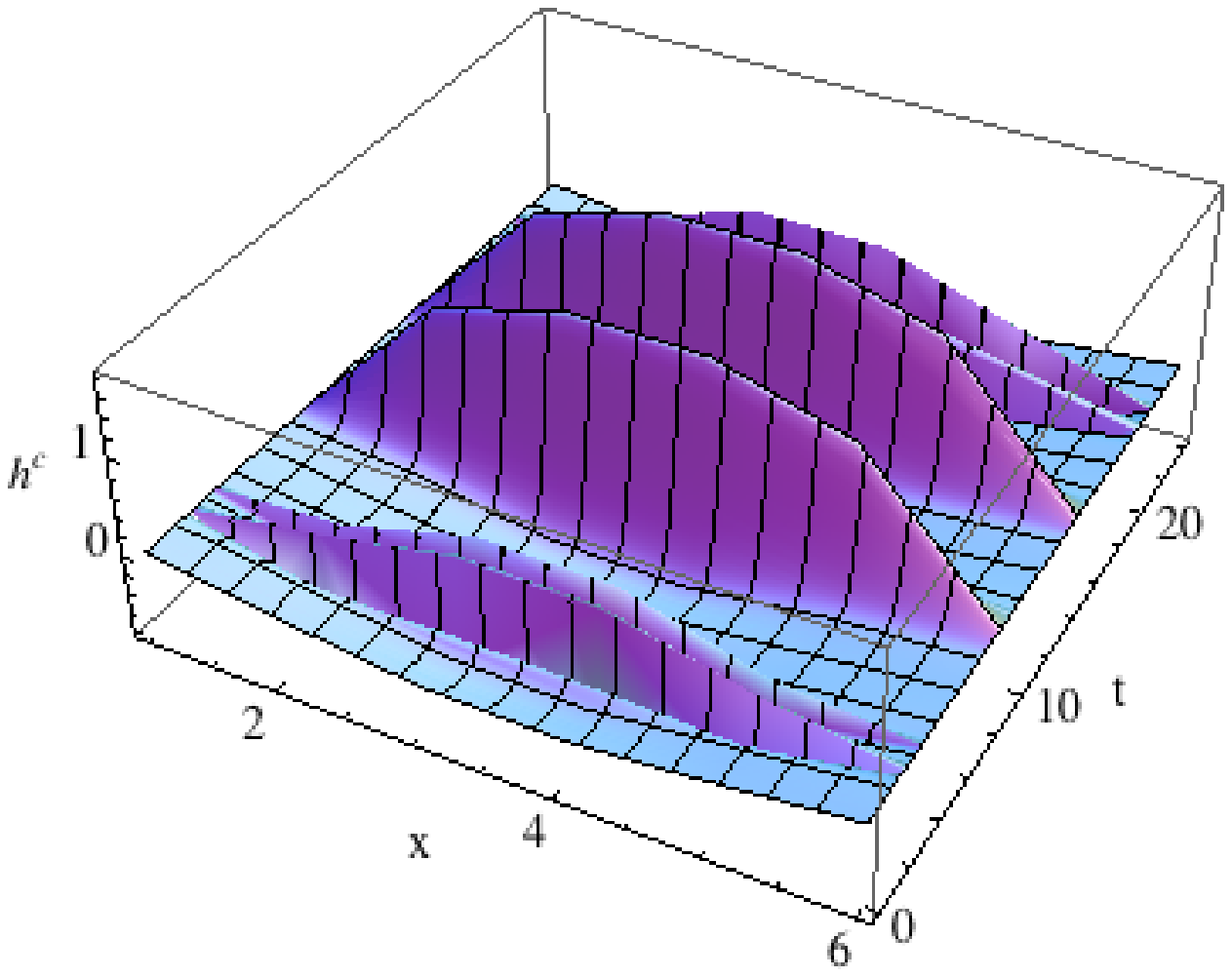}} \hspace{0.1\columnwidth}
\subfigure[]{\includegraphics[width=1.\columnwidth]{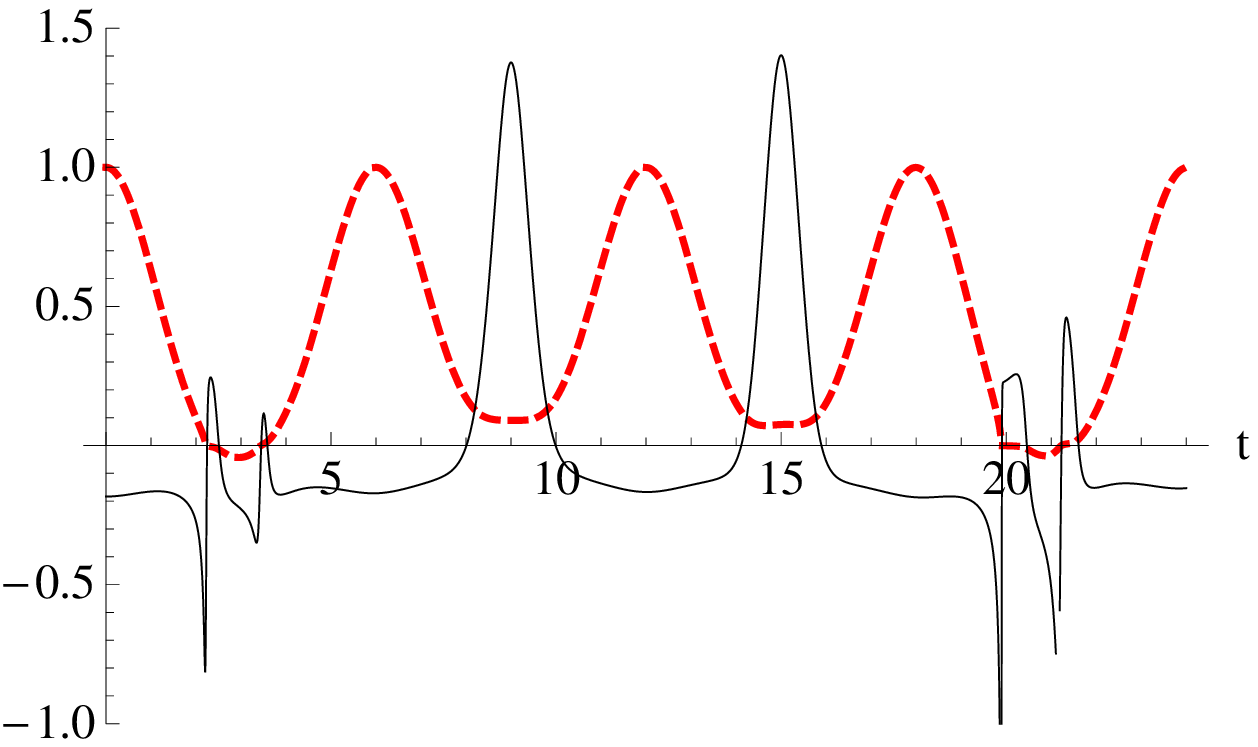}}
\end{array}$
\caption{$s=6$; Frame (a): the magnetization field determined by the target Hamiltonian \eref{eq:hDatta}. Frame (b): the magnetization field determined by the controlled Hamiltonian \eref{eq:hcontrHam}.  Frame (c):
the control field $h^c(x,t)$, obtained by numerical integration of \eref{eq:diffalg}, as a function of $x$ and $t$; interpolation in $x$ is only for graphical convenience. Frame(d): the control field $h^c(2,t)$ (solid line) and the the mean value of $\tau(1,2)$ in the controlled state (dashed line). In order to avoid singularities in $h^c(x,t)$ due to the zeroes of this mean value we have used Tikhonov regularization \cite{baer08}. \label{fig:figHam}}
\vspace{-0.55cm}
\end{figure*}

\begin{figure*}[h]
\centering
$\begin{array}{c}
\subfigure[]{\includegraphics[width=1.\columnwidth]{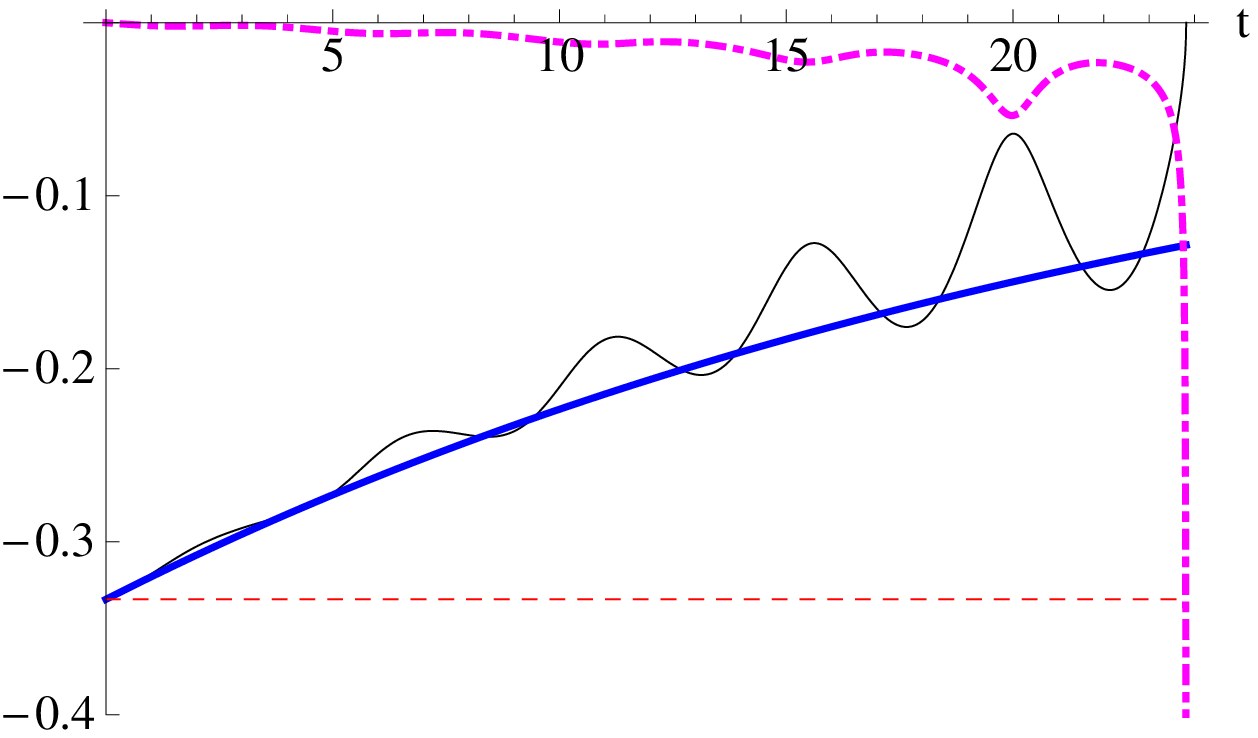}}\hspace{0.\columnwidth}
\subfigure[]{\includegraphics[width=1.\columnwidth]{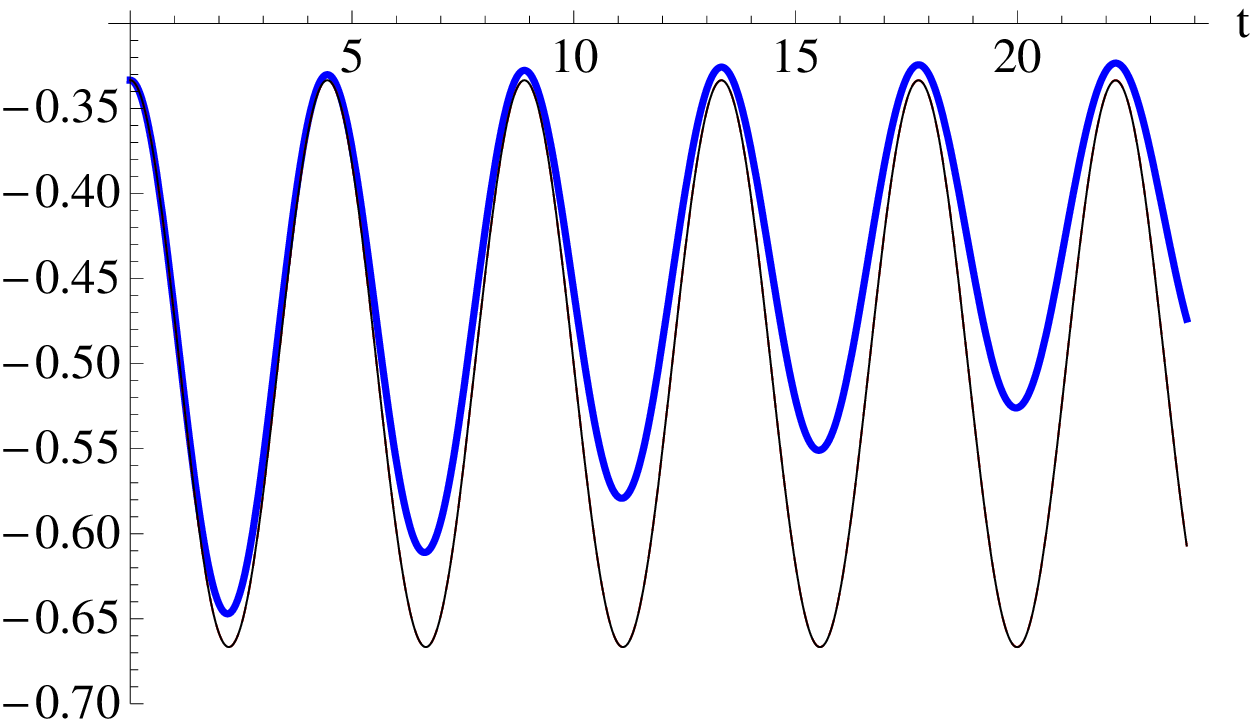}}
\end{array}$
\caption{$s=3,\ \eta=0.01,\ \epsilon=0$. Frame (a): the control field $h^c(2,t)$ is represented (rescaled by a factor 0.1 for graphical convenience) as a function of time as a dot-dashed thick line. Expectations of the local kinetic energy $T(1)$, under three different evolutions,  are represented as functions of time: the dashed line corresponds to pure Hamiltonian; the  solid thick line to Hamiltonian+dephasing; the solid thin line to Hamiltonian+dephasing+control field. Frame (b): the expectations of $\sigma_3(1)$ for the same three systems. Same graphical conventions as in frame (a); the dashed line is not visible since the magnetizations of the target and controlled systems are, as long as both exist, numerically indistinguishable. \label{fig:figDeph}}
\end{figure*}

\begin{figure*}[h]
\centering
$\begin{array}{c}
\subfigure[]{\includegraphics[width=1.\columnwidth]{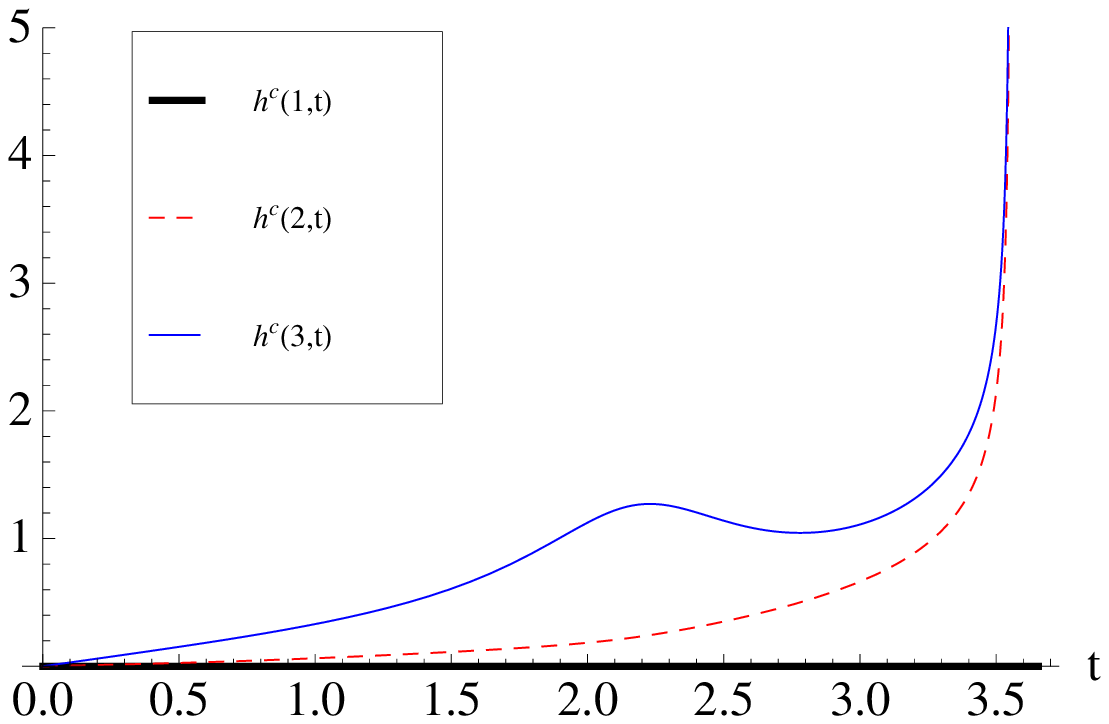}}\hspace{-0.0\columnwidth}
\subfigure[]{\includegraphics[width=1.\columnwidth]{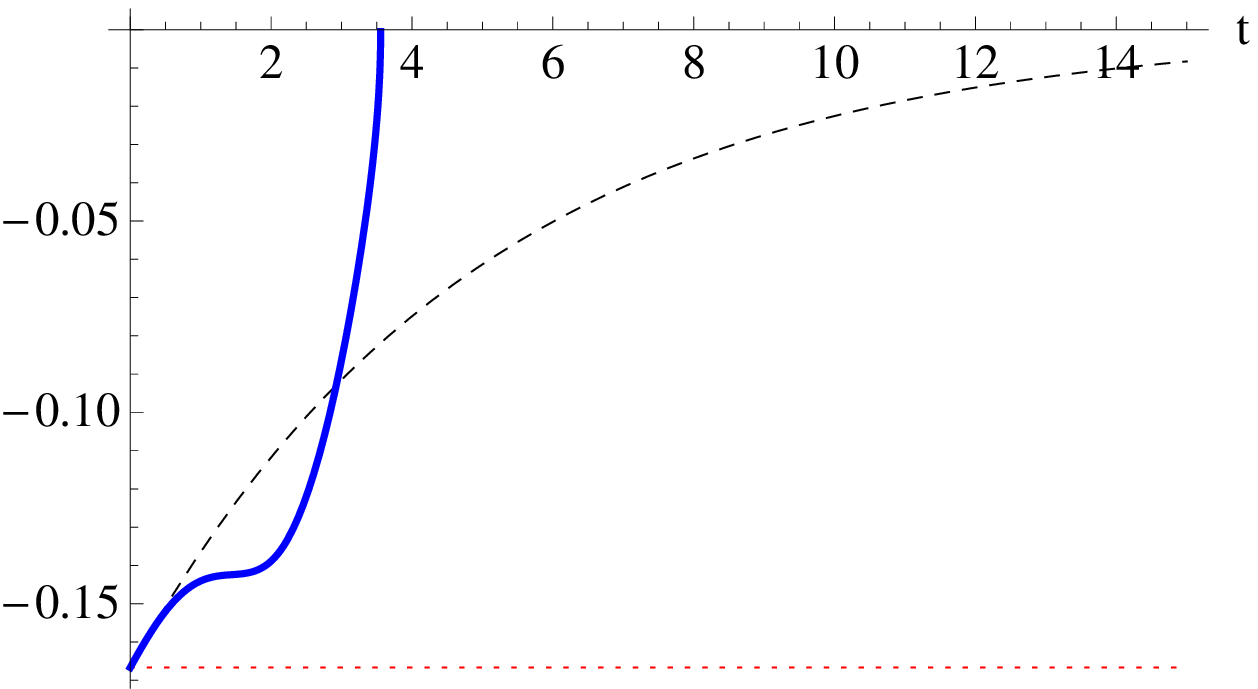}}\\
%\vspace{-0.2cm}\\
\subfigure[]{\includegraphics[width=1.\columnwidth]{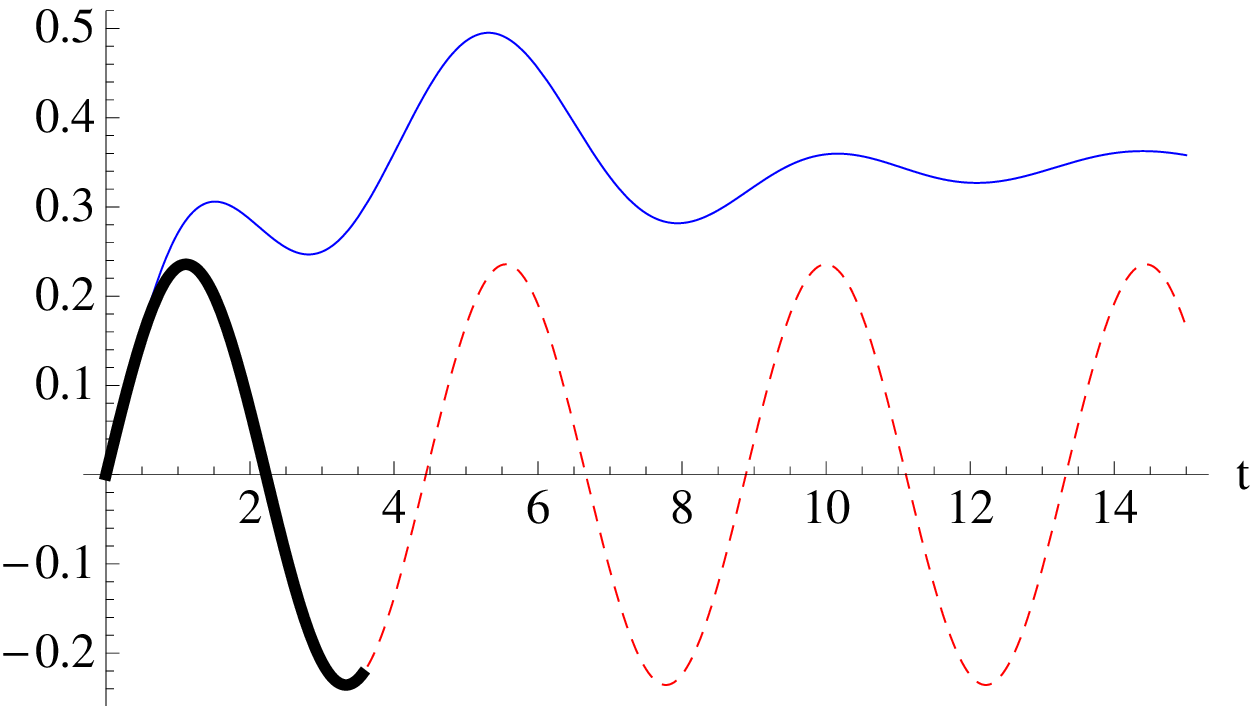}} \hspace{-0.0\columnwidth}
\subfigure[]{\includegraphics[width=1.\columnwidth]{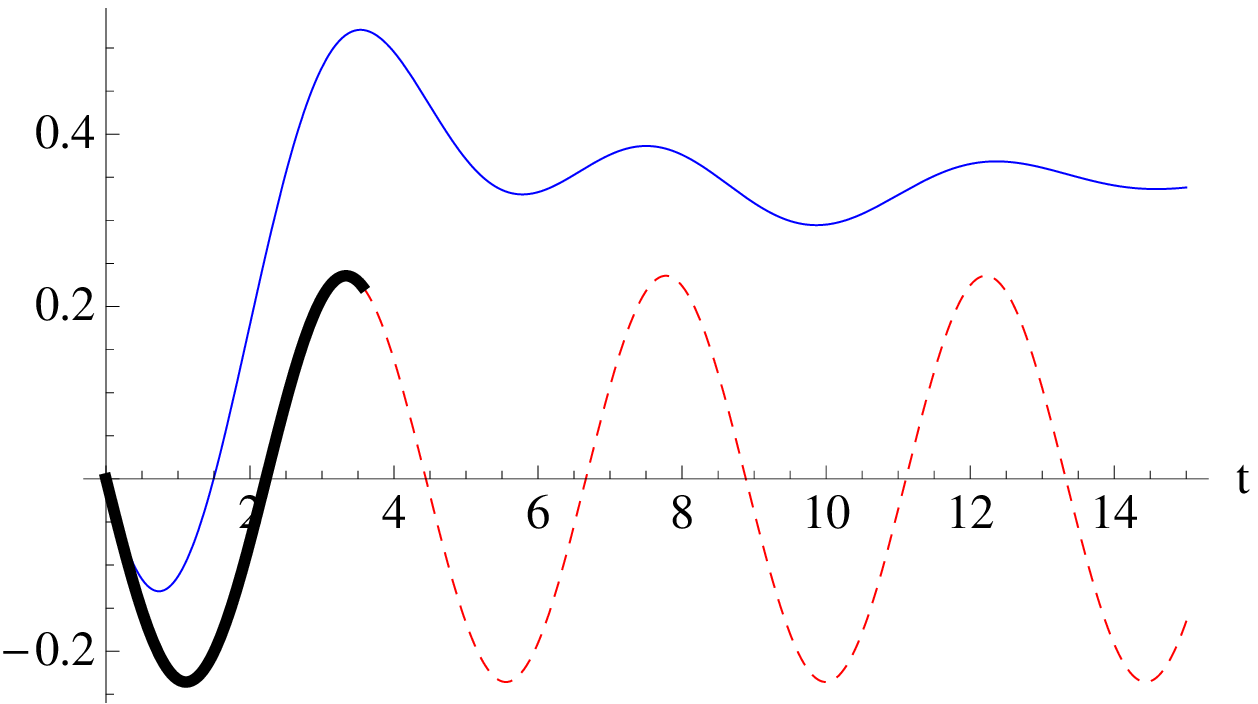}}
\end{array}$
\caption{$s=3,\ \epsilon=0.1,\ \mu=-1,\ \eta=0$. Frame (a): the field $h^c(x,t)$. Frame (b): $Tr\llrr{T(1) \overline{\rho}(t)}$ (dotted line); $Tr\llrr{T(1) \rho^{actual}(t)}$ (dashed line); $Tr\llrr{T(1) \rho^c(t)}$ (solid thick line). Frame (c): $\bar{j}(1,t)$ (dashed line); $j^c(1,t)$ (solid thick line); $j^{actual}(1,t)$ (solid thin line). Frame (d): $\bar{j}(2,t)$ (dashed line); $j^c(2,t)$ (solid thick line); $j^{actual}(2,t)$ (solid thin line). The current field $j^c(x,t)$ is, as long as it exists, numerically indistinguishable from the target field $\overline{j}(x,t)$.\label{fig:figBaths}}
\end{figure*}

\begin{figure*}[h]
\centering
$\begin{array}{c}
%\vspace{-0.5cm}\\
\subfigure[]{\includegraphics[width=1.\columnwidth]{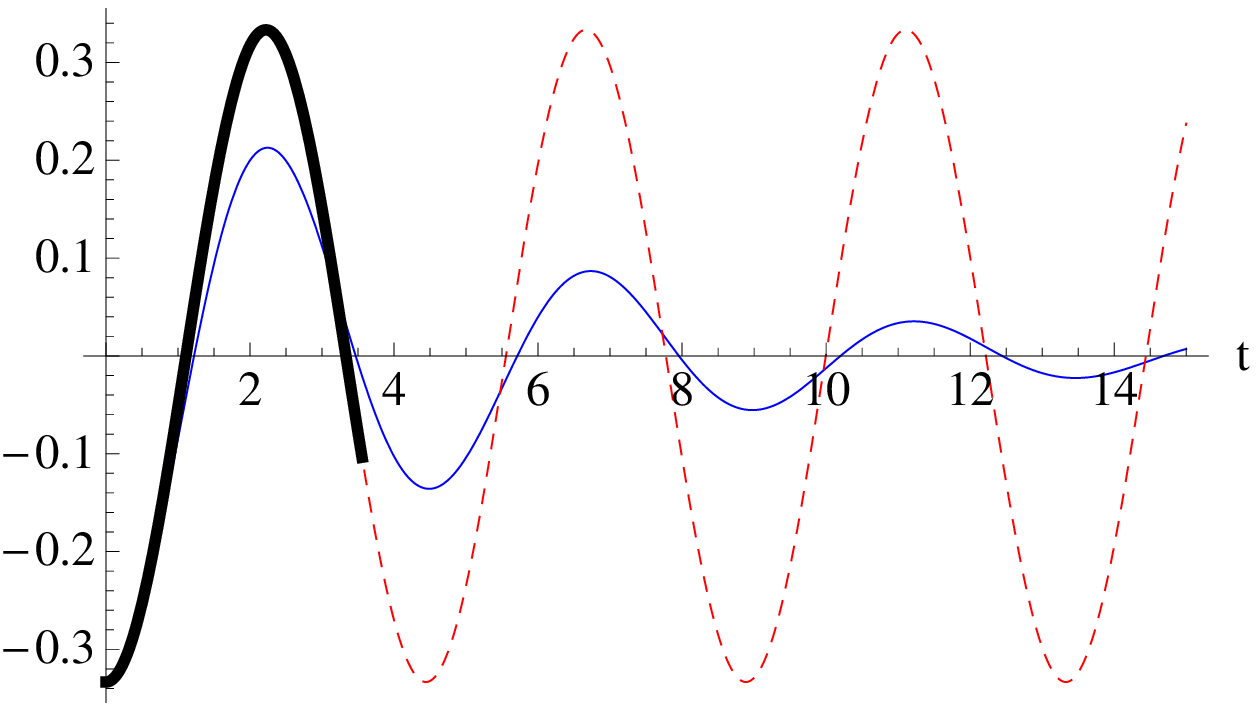}} \hspace{0.\columnwidth}
\subfigure[]{\includegraphics[width=1.\columnwidth]{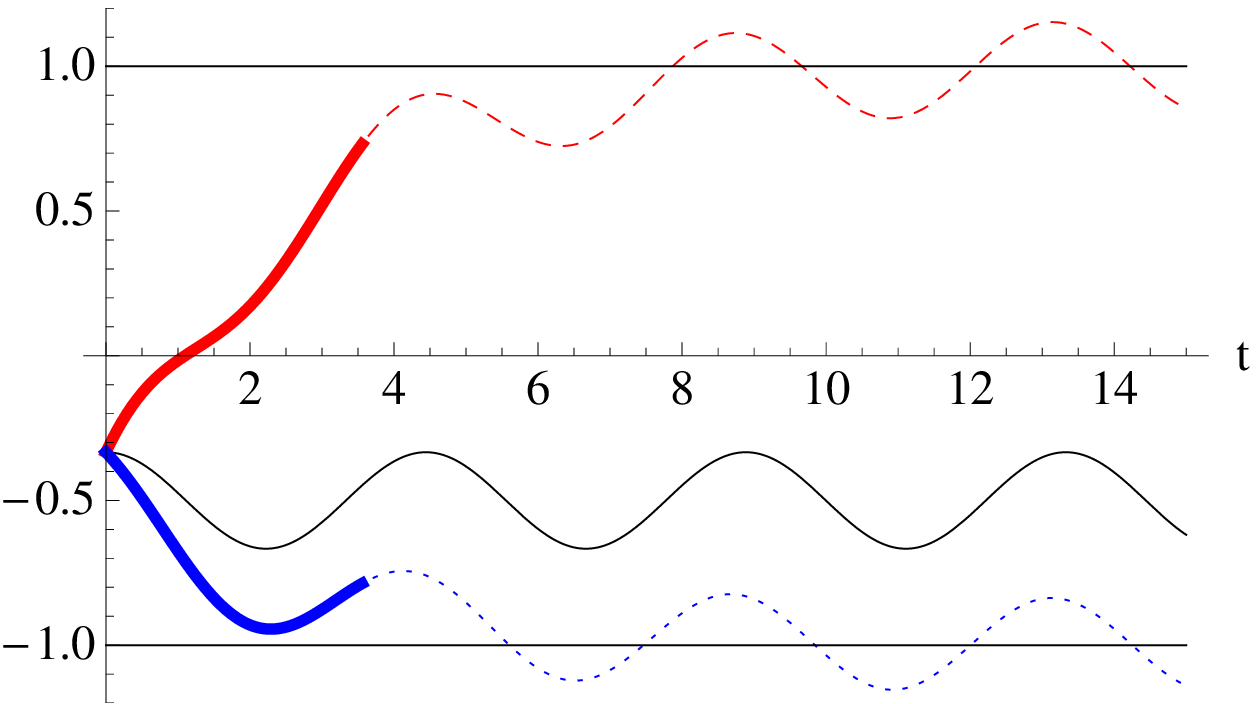}}
\end{array}$
\caption{$s=3,\ \epsilon=0.1,\ \mu=-1,\ \eta=0$. Frame (a): $Tr\llrr{\sigma_3(2) \rho^{actual}(t)}$ (solid thin line); $\overline{m}_3(2,t)$ (dashed line); $m_3^c(2,t)$ (solid thick line). The magnetization $m^c(2,t)$ is, as long as it exists, numerically indistinguishable from the target magnetization $\overline{m}_3(x,t)$. Frame (b): $a(t)$ and $b(t)$ as functions of time (dashed and dotted thin lines, respectively); the corresponding fields $m_3^c(1,t)$ and $m_3^c(3,t)$ determined by numerical (local) solution of the VL problem are represented as thick lines; solid thin line: the target magnetization $\overline{m}_3(1,t)=\overline{m}_3(3,t)$. For the extremal sites of the chain, the controlled system does not attain, even locally, the target magnetization.\label{fig:figBathsMag}}
\end{figure*}

%%%%%%%%%%%%%%%%%%%%%%%%%%%%%%%%%%%%%%%%%
We have probed the interesting idea, advanced in reference \cite{aspuru11}, of \emph{quantum computing}, by means of spin $1/2$ arrays, \emph{without wavefunction}.
\\In our exposition we have stressed the dynamic control perspective advanced in \cite{aspuru11}: the components of the wave function or, in our generalization to the case of open quantum systems, of the density matrix, are the \emph{latent} dynamical variables, that in their evolution (parametrized by $J,\ K,\ \epsilon,\ \mu,\ \eta, \ket{\psi_0}$) determine the response, in terms of \emph{output} magnetization and/or current, to the time-dependent \emph{input} variables $h^c(x,t)$.
\\[5pt]In our choice of numerical examples, we have privileged themes motivated by our previous experience with Feynman's model of a quantum computer \cite{feyn86} \cite{peres85} \cite{nagaj10}.
\\For instance, the numerical example of \fref{fig:figHam} can be rephrased in the following way: can one use the freedom of reparametrizing the system ($J,\ K, \ldots,\ket{\psi_0} \to J',\ K', \ldots, \ket{\psi_0'}$) without altering the \emph{input-output} response in such a way as to simulate, on a transitionally invariant structure, the delicate tailoring of hopping parameters required for perfect transfer of the \emph{cursor} along the \emph{program line}?
\\The example of \fref{fig:figDeph} can be similarly reformulated as: can one compensate by external fields for, in the words of Feynman \cite{feyn86}, \emph{imperfections and free energy losses} in the motion of the \emph{clocking cursor}?
\\[5pt]In answering the above questions, the success of TDDFT has been only partial because the solutions exist only locally in time: the point is that in equations \eref{eq:diffalg} and \eref{eq:algdiffRev} the unknown gradient $h^c(x+1,t) - h^c(x,t)$ appears with the local mean kinetic energy as a coefficient (a \emph{handle}).
\\This difficulty would turn out to be particularly severe if one tried simulate a sharp initial condition in the Hamiltonian context of \fref{fig:figHam}: the \emph{handles} are in this case initially zero, to start with.
\\In the context of \fref{fig:figDeph} (dephasing), vanishing of the \emph{handles} is unavoidable, because they are precisely the nondiagonal matrix elements of the density matrix that dephasing is supposed to damp.
\\The example of \fref{fig:figBaths} and \fref{fig:figBathsMag} (interaction with baths) shows that it is indeed possible to compensate for dissipation even when the number of particles is no longer a constant of motion, at least for the spins that are not coupled with the baths. It is not surprising that the magnetizations of the extremal sites is not attainable by means of the control field: the field $h^c$ does commute with the number operator and therefore  is not able to absorb injected charges or to  create new ones. 
\\[5pt]Further efforts are needed, we think, in the choice and/or in the smoothing of the initial condition and in the introduction of ``convergence factors'' \cite{aspuru11} that extend the existence interval of the solution with little errors in the overall propagation.
%%
%\\[5pt]That a Kohn-Sham potential makes it possible to reproduce the Lindblad dynamics of an open system with a closed, unitarily evolving one, was already known.
%\\What is new is that the same technique can be used, at least in principle, to compensate for the undesirable effect of the environment on a quantum system.
%\\Severe limitations:
%\begin{itemize}
%\item Solutions are only local: is it a numerical problem that can be solved by suitable numerical strategies? Are there physical limitations of some kind?
%\item Suitable choice of the initial conditions to extend the life of the local solution.
%\item Currents are always reproduced, as long as the local solution exists.
%\item If $N_3$ is a constant of  motion, magnetizations can be reproduced for all sites as well, thanks to the continuity equation.
%\item If $N_3$ is not a constant of motion: the magnetization of internal sites $ 1<x<s$ is reproduced as long as the solution exists. Extremal spins: the actual evolution does not conserve the number operator. Since it is impossible to restore $N_3$ to its initial value via the control field, the magnetization of the extremal sites is not reproducible. What can be said is that, from \eref{eq:leftend} it follows trivially that:
%\begin{align}
%  \overline{m}_3(1,t)  &= m_3^c(1,t) +  \\
% & + \mu\llrr{1-e^{-4 \epsilon t}}  +  4 \epsilon e^{-4 \epsilon t} \int_0^t \overline{m}_3(1,\tau) e^{\epsilon t} d\tau.  \nonumber  
%\end{align}
%\\Same reasoning for $m_3(s,t)$.
%\item Online/offline control.
%\end{itemize}
%%
%

%%%BIBLIOGRAPHY
%\bibliography{TDDFT_spin_chains}

\end{document}